\documentclass{elsart}
\usepackage{epsfig}
\usepackage{amssymb}
\renewcommand{\vec}[1]{\mbox{\boldmath $#1 $}}

\begin{document}
\begin{frontmatter}
\title{Combining Hebbian and reinforcement learning in a minibrain model}
\author{R. J. C. Bosman, W. A. van Leeuwen and B. Wemmenhove}
\address{Institute for Theoretical Physics, University of Amsterdam,
Valckenierstraat 65, 1018 XE Amsterdam,  
The Netherlands}

\begin{abstract}
A toy model of a neural network in which both Hebbian learning and 
reinforcement learning occur is studied. The problem of `path interference',
which makes that the neural net quickly forgets previously learned 
input--output relations is tackled by adding a Hebbian term (proportional
to the learning rate $\eta$) to the reinforcement term (proportional to 
$\rho$) in the learning rule. It is shown that the number of learning steps
is reduced considerably if $1/4 < \eta/ \rho < 1/2$, i.e., if the 
Hebbian term is neither too small nor too large compared to the reinforcement
term.
\end{abstract}
\end{frontmatter}

\newpage
\section{Introduction}\label{sec:intro}
The central question which we address in this article is in what way a
biological neural network, i.e., the brain, or, more generally, a part
of the nervous system of an animal or human being, may learn to realize
input--output relations. 
By biological we here mean: realizable with the help of elements
occurring in nature, e.g., neurons or chemical substances that may influence
other neurons or the synaptic efficacy.

An example of an input--output relation is a motor task, like catching
a prey, in reaction to visual, auditive, or other input. 
Many attempts to explain the way input--output relations of this
kind might be realized by (artificial) neural nets are encountered in the 
literature, 
most of which are not satisfactory from a biological
point of view as we will illustrate in subsection \ref{subsec:nonbio}.

It is the purpose of this article to combine ideas which do 
satisfy certain biological
constraints and study a toy model, in particular with respect to
its ability to learn and realize input--output relations.

The widely accepted idea of Hebbian learning
\cite{He49} at the one hand will be combined with some rule that implements 
a feedback signal at
the other hand, in a way that, in principle, might be biologically
realizable. Without the addition of any feedback-signal, learning of
prescribed input--output relations ---whether in reality or in a
model--- is, of course, impossible.

\subsection{Artificial learning rules} \label{subsec:nonbio}
If one wants a network to learn to realize input--output relations,
there are various well-known prescriptions, associated with names like
perceptron learning rule, back-propagation or Boltzmann machines
\cite{Mu90,He91}. None of these, however, model
the functioning of real brains, 
since the learning rules in question violate the existing biological 
limitations. In order to illustrate this statement, let us give an 
example.

Consider a single layered feed-forward network, i.e., a network consisting 
of an input and an output layer only, in which signals 
are sent by neurons of the input layer to neurons of the output layer, and
not the other way around. Let $w_{ij}$ be the strengths or `weights'
of the connections in this simple net. In 1962, Rosenblatt
\cite{Ro62} proved that such a network will realize desired
input--output relations if, a finite number of times, the weights are 
adapted according to the rule
\begin{equation}
w_{ij} \rightarrow w_{ij} + \Delta w_{ij}
\end{equation}
with
\begin{equation}
\Delta w_{ij} = \varepsilon(x_{{\rm{T}}i} - x_{{\rm{O}}i})x_j
\label{eq:percep1}
\end{equation}
where $x_{{\rm{T}}i}$ is the desired or target output of neuron $i$, and
$x_{{\rm{O}}i}$ is its actual output. Furthermore, $x_j$ is the state of 
the pre-synaptic input neuron $j$ and $\varepsilon$ 
is some function of the neuron states and properties of neurons $i$ and $j$.
This learning rule can not be realized by a biological neural net
since neuron $i$, producing $x_{{\rm{O}}i}$, cannot know that it
should produce $x_{{\rm{T}}i}$.
If, e.g., an animal does not succeed in catching a prey, its neurons get
no specific feedback individually, on what the right output $x_{{\rm{T}}i}$ 
should have been.
Hence, $x_{{\rm{T}}i}-x_{{\rm{O}}i}$ cannot be determined by the biological 
system, and,
therefore, neither can it adapt the weights according to (\ref{eq:percep1}). 
Consequently, the perceptron learning rule (\ref{eq:percep1})
is unsuitable for a realistic modeling of the way in which a biological 
neural net can learn and realize input--output relations. Similar 
observations can be made for back-propagation or Boltzmann machines.

\subsection{Biological learning rules; Hebbian learning and reinforcement
learning}
\label{subsec:biohebb}
Already in 1949, Hebb suggested \cite{He49} that, in biological 
systems, learning takes place through the adaptation of the strengths of 
the synaptic interactions between neurons, depending on the activities
of the neurons involved. 
In a model using binary neurons, i.e., $x_i = 0$ or $x_i = 1$, the most 
general 
form of a learning rule based on this principle is a linear function in 
$x_i$ and $x_j$ since $x_i^2=x_i$ and $x_j^2=x_j$. It therefore reads
\begin{equation}
\Delta w_{ij} = a_{ij} + b_{ij}x_i + c_{ij}x_j + d_{ij}x_i x_j
\label{genhebb}
\end{equation}
In a biological setting, the coefficients $a_{ij}, b_{ij}, c_{ij}$ and $d_{ij}$
in this learning rule can only depend on locally available information, 
such as the values of the membrane
potential 
\begin{equation}
h_i = \sum_j w_{ij} x_j
\label{membranepotential}
\end{equation}
and the threshold potential 
$\theta_i$ of neuron $i$. In this
way, the system adapts its weights without making use of neuron-specific
information, like, e.g., $x_{{\rm{T}}i}$, of which there can be no knowledge,
locally, at the position of the synapse.

In a recurrent neural net, a Hebbian learning rule 
suffices to store patterns
\cite{He99,Mu90} if all neurons are clamped to
the patterns which are to be learned during a certain period, the 
`learning stage'. In feed-forward networks, however, only the neurons of the 
input layers are clamped, and some kind
of feed-back signal, governing the direction of adaptation of the 
synapses during the learning procedure, is indispensable. Probably the simplest
form of such a signal one can think of is a `success' or
`failure' signal that is returned to the network after each attempt to
associate the correct output to given input.
On the basis of trial and error, a neural network can then indeed learn 
certain 
tasks, the principle on which
`reinforcement learning' is based \cite{Ba95,Su98}. 
This principle of reinforcement learning has a rather natural interpretation:
satisfactory behavior is rewarded,
or \emph{reinforced}, causing this behavior to occur more frequently. 
The reinforcement signal is supplied by the subject's environment, or by its
own judgment of the effect of its behavior.
In a biological perspective, one could think of the synaptic 
change being influenced by some chemical substance, which is released depending
on whether the evaluation by the subject of the effect of the output is 
positive or negative, i.e., whether it is happy or unhappy with the
attempt it made.

Note that, in learning by reinforcement,
the search for the correct output is more difficult, and, hence, 
slower, than for non-biologically realizable 
algorithms like the perceptron learning rule or back-propagation. This is not 
surprising, 
since the latter give the system \emph{locally} specific information on how to 
adjust individual weights, while
reinforcement rules only depend upon a \emph{global} `measure of correctness'.

The most general form of a reinforcement learning algorithm is given by the
prescription
\begin{equation}
\Delta w_{ij} = a_{ij}(r) + b_{ij}(r)x_i + c_{ij}(r)x_j + d_{ij}(r)x_i x_j
\label{genreinf}
\end{equation}
Here the coefficients $a_{ij},b_{ij},c_{ij},d_{ij}$, besides their dependence 
on the local variables such
as the membrane potential, will in principle also depend on a reinforcement
signal, denoted by $r$. The value of $r$ is usually a real number between
$0$ and $1$, denoting the degree of success ($r=1$ means success, $r=0$ means
failure).

An important issue in the literature on reinforcement learning is the so 
called `credit assignment problem' \cite{Su98}. 
It refers to the question how a neural
net knows which connections $w_{ij}$ were responsible for a successful or
unsuccessful trial, and, as a consequence, which connections should be
`rewarded', and which should be `punished', respectively. 

In their article `Learning from mistakes' (1999),
Chialvo and Bak \cite{Ch99},
proposed a class of networks, in which learning takes place 
on the basis of a `deinforcement signal' only, i.e., the weights of
active synapses are decreased if the output
is wrong, they are `punished', so to say, in case of wrong performance
of the network. If the output is right nothing happens. This procedure
works as long as the average activity in the net is kept very low:
when only a few neurons are active 
at an unsuccessful attempt, one can be sure that the connections between 
these active neurons were 
the ones which were responsible, and thus should be `punished'.
In this way Chialvo and Bak obtained an elegant solution to the credit 
assignment problem.

The absence of a reinforcement signal (nothing happens if $r=1$)
makes their learning rule relatively simple. It is a version of the general
rule (\ref{genreinf}) with $b_{ij} = 0$ and $c_{ij}=0$:
\begin{equation}\label{eq:Bak1}
\Delta w_{ij} = -(1-r) (\rho x_{i}x_{j} - \varphi)
\end{equation}
where $\rho$ and $\varphi$ are positive constants; in this article we will
suppose $\varphi << \rho$.
A biological mechanism that could implement the learning rule 
(\ref{eq:Bak1}) is the 
following: if the output is correct, nothing happens, since the network 
obviously performs satisfactory. If not, a chemical substance is released, 
which has the effect that synapses between neurons that have just been active,
and thereby are `tagged' in some electro-chemical way, are depressed. 

\subsection{Purpose} \label{subsec:outline}

The success of the `minibrain model' of Chialvo and Bak \cite{Ch99}
(as Wakeling and Bak 
referred to it in \cite{Wa01}), is limited to 
feed-forward neural nets in which the number of input and output neurons
(or, equivalently in this model, the number of patterns) is small compared to 
the number of
neurons in the hidden layer. As the number of neurons in the hidden
layer decreases, learning, at a certain moment, becomes impossible:
`path interference' is the phenomenon which causes this effect \cite{Wa02}.
Essentially, it amounts to the following.
If, in each layer of the feed-forward neural net, only one neuron is active
at each time step, an input--output relation corresponds to a path of activity
along the strongest connections between the neurons.
Basically, path interference comes down to the erasure of an
existing path of activity, which was correct at a previous learning step,
by a change due to a punishment of a connection while trying to learn a 
different input--output relation. 
If the probability for this path interference to occur becomes too
large, learning times tend to infinity.

In this article we attempt to improve the performance of the minibrain model
of Chialvo and Bak
---in the sense of decreasing the learning time--- by making sure that, at the
occurrence of path interference, the punishment of a correct activity path is 
no longer such that the memory is erased.
We achieve this by adding to the deinforcement term in the learning rule 
(\ref{eq:Bak1}), which is proportional to $\rho$, 
a Hebbian term proportional to $\eta$. The latter term has the 
effect that an active path is strengthened, mitigating in this way the 
punishment.
By choosing the ratio between the coefficients $\eta$ and $\rho$ of both 
terms in the 
learning rule appropriately, we are able to reduce the number of learning 
steps significantly, without making the model less realistic from a biological
point of view. In fact, in the class of 
models we study, Hebbian learning is a most appropriate way to account 
for biological observations like LTP and LTD \cite{Ka91}. 
In section \ref{sec:expl} we explain that if the quotient of 
the Hebbian learning rate and the coefficient of the deinforcement term
is in the range
\begin{equation}
\frac{1}{4} < \frac{\eta}{\rho} < \frac{1}{2}
\label{ineq}
\end{equation}
learning times are reduced considerably.

In their article \cite{Ba01}, Chialvo and Bak proposed a different way to 
solve the problem of path interference. They reduced the amount of 
punishment of the connections that once had been active in a correct 
response. In this model a neuron needs to memorize whether it 
previously was active in a successful trial. In our model such a neuron 
memory is not needed.

Let us denote the deinforcement contribution to learning by $\Delta w_{ij}'$ 
and denote the Hebbian part by $\Delta
w_{ij}''$. We will study a learning rule of the form
\begin{equation}
\Delta w_{ij} = \Delta w_{ij}' + \Delta w_{ij}''
\label{eq:Totalrule1}
\end{equation}
From all possibilities for
Hebbian learning summarized by equation 
(\ref{genhebb}), we choose for $\Delta w_{ij}''$ a rule in which
the coefficients $a_{ij}$ and $b_{ij}$ both are zero:
\begin{equation}
\Delta w_{ij}'' = \varepsilon(x_i, x_j)(2x_i-1)x_j
\label{hebbrule}
\end{equation}
We choose this particular form since it can be argued that this form is a most
plausible candidate from a biological point of view \cite{He99}.

Our paper has been set up as follows. 
In section \ref{sec:model}, we describe a feed-forward network with
one hidden layer, which we will
use to study the learning rule (\ref{eq:Totalrule1}), with (\ref{eq:Bak1}) and
(\ref{hebbrule}). 
In section \ref{sec:numsim}, 
numerical studies of various situations are given and explained. It
turns out, in general, that taking into account Hebbian learning,
and viewing it as a process which is permanently active, 
irrespective of the occurrence 
of reinforcement learning, has a
positive influence on the learning time of the neural net. 
This is a new result, 
which, to the best of our knowledge, has not been noticed
before.
\section{Description of the model: updating rules for neuron activities and 
connection weights}\label{sec:model}
\begin{figure}
\begin{center}
\input{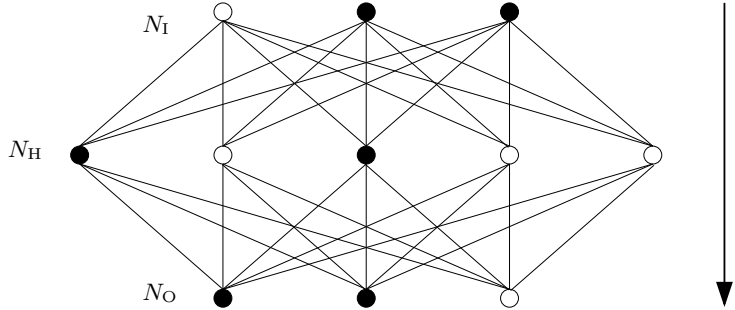}
\caption{An example of a fully connected feed forward network with $N_{\rm{I}}$
input neurons, $N_{\rm{H}}$ hidden neurons and $N_{\rm{O}}$ output
neurons. The filled circles represent active neurons.}
\label{fig:model}
\end{center}
\end{figure}

In order to explore a simplified model of the brain we
consider a fully connected, feed-forward neural network with an input
layer of
$N_{\rm{I}}$ neurons, one hidden layer of $N_{\rm{H}}$ neurons, and an
output layer of $N_{\rm{O}}$ neurons, see figure \ref{fig:model}.
The state $x_{i}$ of neuron
$i$ is $1$ if neuron $i$ fires and $0$ if it does not.
In general, a neuron fires if its
potential $h_{i}$ is sufficiently high, where $h_{i}$ stands for the
membrane potential $V_{\rm{ex}}-V_{\rm{in}}$, the difference
between the intra- and extra cellular potentials $V_{\rm{in}}$ and
$V_{\rm{ex}}$.
Following Chialvo and Bak \cite{Ch99}, we model the dynamics of the neural net
by simply stating that in the hidden and 
output layers a \emph{given} number of neurons having
the highest potentials  $h_{i}$ ---in their respective layers--- are
those that will fire. In their terminology: we use \emph{extremal dynamics}. 

For McCulloch and Pitts
neurons a way to control the average activity might be realized via 
suitably chosen 
threshold potentials $\vartheta_{i}$
 (see e.g.  \cite{Al95}, \cite{Ba95}). 
In nature, the average activity will depend on the threshold potentials and
may, moreover, be influenced
by chemical substances or the network structure \cite{Ka91,Ko90}.
In our model we put the threshold potentials $\theta_i$ equal to zero:
\begin{equation}\label{thetazero}
\theta_i = 0
\end{equation}

The number of neurons in the input, hidden and output layers that we 
choose to be \emph{active},
will be denoted by $N_{\rm{I}}^{\rm{(a)}}$, $N_{\rm{H}}^{\rm{(a)}}$ and
$N_{\rm{O}}^{\rm{(a)}}$, respectively.

The input pattern, a specific set of states of neurons in the input layer, 
will be denoted by 
$\vec{\xi}_{\rm{I}}=(\xi_{{\rm{I}}1},...,\xi_{{\rm{I}}N_{\rm{I}}})$. 
The network
is to associate every input pattern with a desired, or target,
output pattern, $\vec{\xi}_{\rm{T}}=(\xi_{{\rm{T}}
1},...,\xi_{{\rm{T}} N_{\rm{O}}})$. The $\vec{\xi}_{\rm{I}}$ and 
$\vec{\xi}_{\rm{T}}$ are vectors with components equal to $0$
or $1$.
Consequently, the number of active neurons of the input and output
layer are given by
\begin{eqnarray}
N_{\rm{I}}^{\rm{(a)}} &=& \sum_{j=1}^{N_{\rm{I}}} \xi_{{\rm{I}} j}\\
N_{\rm{O}}^{\rm{(a)}} &=& \sum_{j=1}^{N_{\rm{O}}} \xi_{{\rm{T}} j}
\end{eqnarray}
In our numerical experiments, these numbers will be taken to be equal.
Moreover, the number of active neurons in the 
hidden layer, 
\begin{equation}
N_{\rm{H}}^{\rm{(a)}} = \sum_{j=1}^{N_{\rm{H}}} x_{{\rm{H}}j}
\end{equation}
where $x_{{\rm{H}}j}$ is the neuron state of neuron $j$ in the hidden layer, 
will also be equal to the number of active neurons in the other layers. 
Hence, we \emph{choose}
$ N_{\rm{I}}^{\rm{(a)}}=N_{\rm{O}}^{\rm{(a)}}=N_{\rm{H}}^{\rm{(a)}}$.

We thus have explicitated the network dynamics. We now come to the update 
rules for the synaptic weights $w_{ij}$. 
Updating of the network state will happen at discrete time steps. At every 
time step $t_n$, all neuron states are updated in the order: input layer -- 
hidden layer -- output layer. This being done,
the values of the weights are updated, according to
\begin{equation}\label{eq:change1}
w_{ij}(t_{n+1})=w_{ij}(t_{n})+\Delta w_{ij}(t_{n})
\end{equation}
Substituting 
(\ref{hebbrule}) and (\ref{eq:Bak1}) into (\ref{eq:Totalrule1}), we find
\begin{equation}\label{eq:weights2}
\Delta w_{ij} = \varepsilon(x_{i}, x_{j})(2x_{i}-1)x_{j} - (1-r)
(\rho x_{i}x_{j} - \varphi)
\end{equation}
For the pre-factor of the Hebbian term we take \cite{He99}
\begin{equation}\label{eq:vareps1}
\varepsilon(x_{i}, x_{j})=\eta\big(\kappa - (2x_{i}-1)(h_{i}-\theta_i)\big)
\end{equation}
The constants $\eta$ and $\kappa$ are positive numbers, the so-called 
learning rate and margin parameter.
Finally, combining the above ingredients and noting that we 
chose $\theta_i=0$, the learning rule reads:
\begin{eqnarray}\label{eq:weights3}
\Delta w_{ij}(t_{n})&=&\eta[\kappa - h_{i}(t_n)(2x_{i}(t_{n})-1)][2x_{i}(t_{n}) -
1] x_{j}(t_{n})\nonumber\\
& & + (1-r)[-\rho x_{i}(t_{n})x_{j}(t_{n}) + \varphi]
\end{eqnarray}
Note that $x_{i}(t_{n})$ and $x_{j}(t_{n})$ are the activities of neurons
in \emph{adjacent} layers, since in our model there are no lateral connections.
The constant $\varphi$ is chosen in such a way that the change in 
$\sum_{i,j} w_{ij}$, where the sum is extended over $i$ and $j$ in 
adjacent layers, due to the $\rho$-term (not due to the Hebbian term), is 
independent of $\rho$. This can easily be achieved by choosing 
$\varphi = \rho/P$, where $P$ is the product of the numbers of neurons in 
two adjacent layers, i.e., $\varphi$ is equal to either 
$\rho/N_{\rm{I}}N_{\rm{H}}$ or $\rho/N_{\rm{H}}N_{\rm{O}}$.

\section{Numerical Simulations}\label{sec:numsim}

The network described in the previous section will now first be studied
numerically. The numerical details are:
\begin{itemize}
\item[--] The initial weights $w_{ij}(t_{0})$ are randomly chosen with
values between $-0.01$ and $+0.01$.
\item[--] The punishment rate $\rho$ will be kept constant at
$0.02$. Thus when we vary the $\eta/\rho$ ratio, we vary the learning
rate $\eta$.
\item[--] The margin parameter $\kappa$, appearing in
(\ref{eq:vareps1}), will be kept at the value $1$.
\item[--] Whenever the number of neurons in the input, hidden or output 
layer is fixed, we choose $N_{\rm I}=8$, $N_{\rm H}=512$ and
$N_{\rm O}=8$.
\item[--] All data are averaged over $512$ samples.
\end{itemize}

The network is confronted with $p$ different input patterns
$\vec{\xi}^{\mu}_{\rm{I}}$, $(\mu=1,\ldots,p)$, to which 
correspond equally many target patterns $\vec{\xi}^{\mu}_{\rm{T}}$.
Learning proceeds as follows. The input layer remains clamped to the first
input pattern until the time step at which the target pattern has been found.
As soon as this input-output relation
$\mu=1$ has been learned, we switch to input pattern $\mu=2$. After the
corresponding target
pattern has been found we continue, up to the $p$-th input-output
relation. Then, we have completed what we will refer to as one `learning 
cycle'.

After this cycle we start the process again, up to the point where the
network can recall all $p$ input-target relations {\it at
once}. When that is the case, learning stops. We count the number of 
learning steps needed to learn all input--output relations. 

Chialvo and Bak consider the case of one active neuron in each layer. 
In section \ref{subsec:actexp} we present a numerical experiment with a neural
network for which the activities are larger than one, i.e., 
$N_{\rm{I}}^{\rm{(a)}} >1$, $N_{\rm{H}}^{\rm{(a)}} >1$ and
$N_{\rm{O}}^{\rm{(a)}} >1$. In particular we study the total number of 
learning steps as a function of the ratio $\eta/\rho$. 
In section \ref{subsec:scale} we vary the number of neurons in
input and output layer and keep the hidden layer fixed, and vice versa. 
Finally, in section \ref{sec:expl}, we interpret
our results.

\subsection{Effect of the Hebbian term}
\label{subsec:actexp}
In `Learning from mistakes' Chialvo and Bak \cite{Ch99} studied the case of
\emph{one} active neuron in the input, the hidden and the output layers:
$N_{\rm{I}}^{\rm{(a)}}=N_{\rm{H}}^{\rm{(a)}}=N_{\rm{O}}^{\rm{(a)}}=1$.
We here will study what happens when $N_{\rm{I}}^{\rm{(a)}}$, 
$N_{\rm{H}}^{\rm{(a)}}$ and $N_{\rm{O}}^{\rm{(a)}}$ all are larger than one.

\begin{figure}[ht]
\begin{center}
\epsfig{file=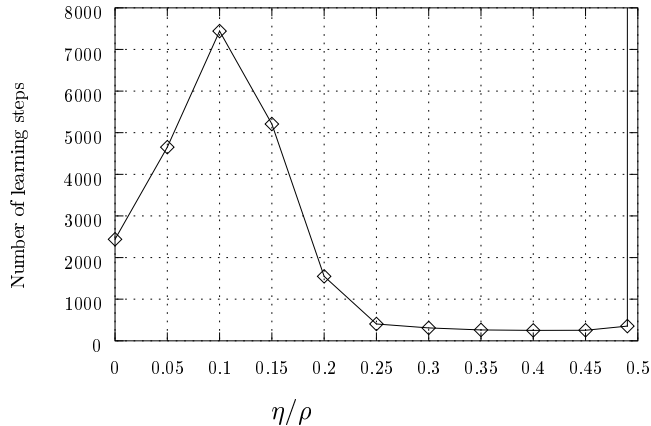, width=0.7\textwidth}
\caption{The number of learning steps as a function of the quotient
$\eta/\rho$. There are eight input--target relations to be learned
and two active neurons in each layer. 
The number of neurons in the input, hidden and output layers are 
$N_{\rm{I}}= 8$, $N_{\rm{H}}=512$ and $N_{\rm{O}}=8$. 
Initially, the number of learning steps increases as a result of the Hebbian 
learning term, but at $\eta/\rho=0.1$ the number of learning steps starts to 
decrease dramatically from $7500$ to $250$ at $\eta/\rho=0.25$. For
$\eta/\rho>0.50$, learning is impossible.}
\label{fig:vareta}
\end{center}
\end{figure}

In our first numerical experiment we take a network with $p=8$ 
input--target relations for which, in each input or target pattern $\mu$, 
the number of active neurons is $2$, i.e., 
$N_{\rm{I}}^{\rm{(a)}}= N_{\rm{H}}^{\rm{(a)}}=N_{\rm{O}}^{\rm{(a)}}=2$. 
In figure \ref{fig:vareta} the number of learning steps is plotted 
against the ratio $\eta/\rho$ of the two proportionality coefficients 
related to the
Hebbian and the deinforcement term respectively.

From figure \ref{fig:vareta} we see that, when there is no Hebbian 
component in the learning rule ($\eta = 0$), the net needs $2500$ learning
steps to learn an input-output task. When we add a slight Hebbian component
($\eta$ small) the number of learning steps increases, and, hence, the ability 
of the net to learn diminishes. However, when the Hebbian component becomes 
more and more important, the number of learning steps starts to decrease 
dramatically: for $\eta/ \rho$ between 
$0.25$ and $0.5$ the number of learning steps is approximately $250$. 
The Hebbian component, which has the 
tendency to engrave previously learned patterns, seems to help to \emph{not} 
forget the old patterns.
If $\eta / \rho$ exceeds the value $0.5$, learning fails. Apparently, the 
`progressive' $\rho$ term, the power of which is to help the network to 
quickly adapt itself to new input-output relations, cannot conquer the 
`conservative' power of the $\eta$-dependent (i.e., the Hebbian) term.
We will come back to these points and consider the effects of the $\eta$ and
$\rho$ terms in some detail in section \ref{sec:expl}.

\subsection{Size dependences}\label{subsec:scale}
In this section we consider the network of figure \ref{fig:model} for 
varying numbers $N_{\rm{I}}=N_{\rm{O}}$ and $N_{\rm{H}}$.

\subsubsection{Effect of varying the sizes of the input and output layers}
In this subsection we test the performance of the network
for various sizes of the input, output and hidden layers. In
particular, we chose to study the size-dependence for three different values
of the learning parameter: $\eta/\rho=0$, $\eta/\rho=0.10$
and $\eta/\rho=0.45$, values which we selected on the basis of the results 
of the previous subsection.

First, we take a network with the fixed number of $512$ neurons in 
the hidden layer, and only one active neuron per layer.
The input and output layers will consist of increasing, equal numbers of 
neurons, starting with $N_{\rm{I}} = N_{\rm{O}} = 4$. 
Moreover, we choose the number of input--output relations $p$ to be 
learned equal to the number of neurons in the input and output layers. 
The input and output layers will be enlarged in steps of $4$ neurons, up to
$N_{\rm{I}}=N_{\rm{O}}=28$ neurons.

\begin{figure}[ht]
\begin{center}
\epsfig{file=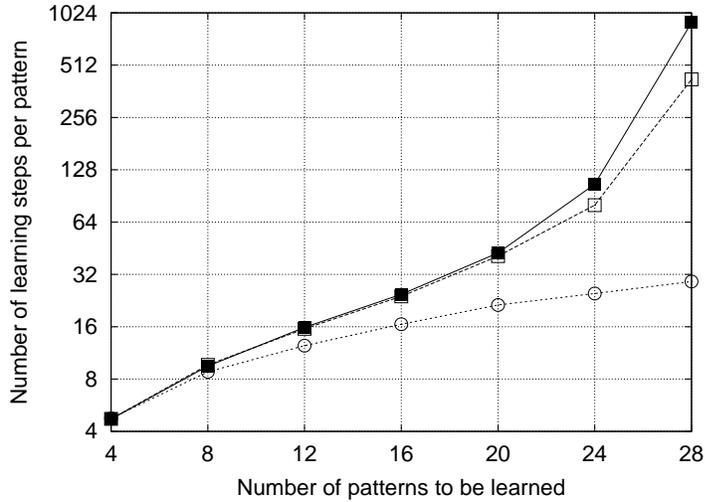, width=0.7\textwidth}
\caption{The number of learning steps per pattern as a function of
the number of input--output relations $p$, for $\eta /\rho =0$ 
($\blacksquare $), $\eta /\rho =0.1$ ($\square$) and $\eta /\rho =0.45$
($\bigcirc$).
Input and output patterns have only one active neuron.
The number of neurons in input and output layers equals the number of
patterns $p$.
Note the logarithmic scale of the  vertical axis. For a small
number of input--output patterns, the learning time is roughly equal for
different values of $\eta/\rho$. The advantageous effect of a Hebbian term 
in the learning rule for this learning task becomes more and more pronounced
with increasing numbers of input--target relations.}
\label{fig:vario}
\end{center}
\end{figure}

In Figure \ref{fig:vario} we give the number of learning steps per
pattern for the above mentioned three values of $\eta/\rho$. 
The positive effect of the addition of a Hebbian term to the learning rule 
becomes more and more convincing with increasing number of input--output 
relations to be learned.

\subsubsection{Effect of varying the size of the hidden layer}

Next we consider a 
network with $8$ input neurons, $8$ output neurons and
$8$ subsequent patterns. The number of active neurons is $2$
for all input and target patterns.
We vary the number of neurons in the hidden layer.

\begin{figure}[ht]
\begin{center}
\epsfig{file=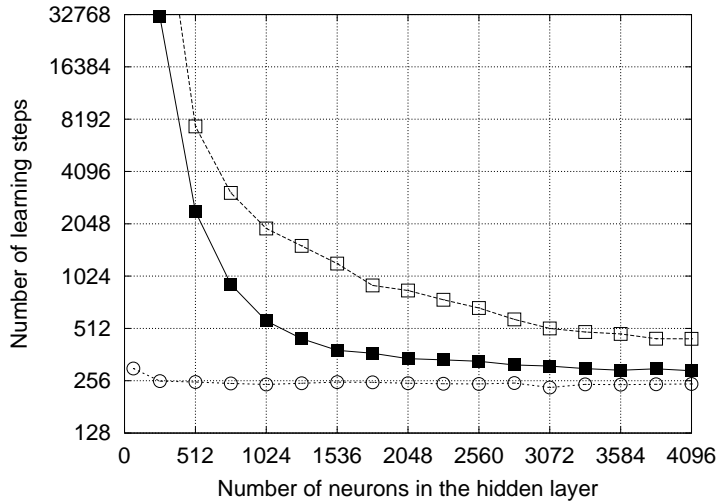, width=0.7\textwidth}
\caption{Dependence of the number of learning steps on the number of neurons 
in the hidden 
layer of the network. The symbols ($\blacksquare$), ($\square$) and 
($\bigcirc$) correspond to
$\eta /\rho=0$, $\eta /\rho=0.1$ and $\eta /\rho=0.45$ respectively.
All input patterns and output patterns
have $2$ active neurons. The number of input neurons, output neurons and 
patterns are fixed; $N_{\rm{I}}=8$, $N_{\rm{O}}=8$, $p=8$.}
\label{fig:varhidden}
\end{center}
\end{figure}

In Figure \ref{fig:varhidden} we have plotted the 
number of learning steps as a function of the number of neurons in the hidden 
layer for three values of the quotient $\eta/\rho$. Note that, in agreement
with figure \ref{fig:vareta} the number of learning steps is the largest
for $\eta/\rho=0.1$ (the symbols $\square$ in figure \ref{fig:varhidden}).
A suitably chosen value for the coefficient $\eta$ of the Hebbian term makes
it possible for the network to perform satisfactory with very small number
of neurons in the hidden layer (the symbols $\bigcirc$ in the figure).

\section{Explanation of the effect of the Hebbian term}\label{sec:expl}
The different behavior for different values
of $\eta/\rho$ is mainly due
to its consequences for the effect we call \emph{path interference}, after 
Wakeling \cite{Wa02}, who studied the critical behavior of the Chialvo \& Bak
minibrain.

As an example, let us consider the case in which only one neuron is active 
in each layer. In this case, the `path of activity' from the active input 
neuron to the corresponding output neuron runs along the outgoing connections 
with the largest weights. 
During the learning process, it is possible that an established path 
(connecting, e.g., the active neuron of pattern $\vec{\xi}_{\rm{I}}^1$
with the active neuron, in the output layer, of $\vec{\xi}_{\rm{O}}^1$) is
`wiped out' by an attempt to learn one of the other input--output relations.
This is likely to happen if the same neuron in the hidden layer 
becomes active, and, consequently, the connection to the output neuron
corresponding to the previously learned pattern is punished by an amount 
$\rho$.
This phenomenon of path interference will occur once in a while,
irrespective of the values of
the parameters $\eta$ and $\rho$. 
However, the question whether the memory of the old pattern
is wiped out (i.e., whether the connection to the output neuron under 
consideration 
is no longer the largest), does depend on the parameters $\rho$ and $\eta$.
To find out how, we should consider the change of this connection 
compared to the change of
the other connections from the same hidden neuron to the other 
output neurons.  
For the total relative change, two different learning steps should be taken 
into account.
Firstly, the one at $t_p$, at which the right output was 
found, and
the deinforcement term did not contribute, and, secondly, the learning
step at $t_q$, at which path interference occurred, and the 
deinforcement
term did contribute.

\begin{figure}
\begin{center}
\epsfig{file=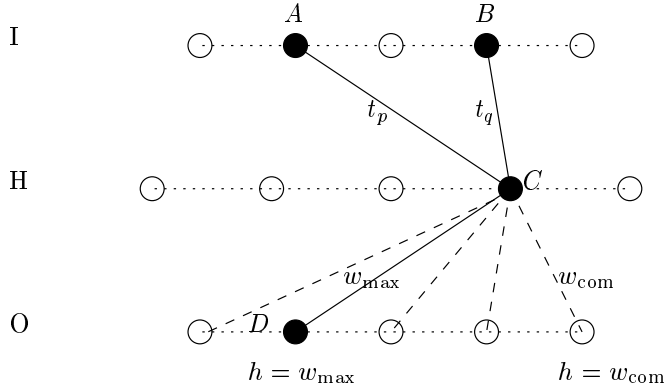}
\caption{\emph{Path interference} At both the time steps $t_p$ and $t_q$ the 
neuron $C$ of the hidden layer fires, and as a result the same neuron $D$
of the output layer is activated. This unwanted effect is due to the fact
that the connection of $B$ and $C$ happens to be the largest one. The paths
$ACD$ and $BCD$ partially overlap.}
\end{center}
\end{figure}

Let $w_{\rm{win}}$ be the largest outgoing weight from the active neuron in 
the hidden layer to the output layer, and let $w_{\rm{com}}$ be a weight
value which is representative for one of the other, competing weights
connecting the same neuron in the hidden layer to a different neuron in the
output layer. 

The membrane potentials
of each neurons $i$ of the output layer are given, according to
(\ref{membranepotential}), by $h_i = w_{ij}$, where $j$ is a neuron of the
hidden layer. 
From (\ref{eq:weights3}), with $x_j=1$, we find in case of success
($r=1$) for the changes of the connections to the winning ($x_i$) and the
competing ($x_i=0$) neurons in the output layer:
\begin{eqnarray}
\Delta w_{\rm{win}}(t_p) & = & \eta[\kappa - w_{\rm{win}}(t_p)] \label{B1} \\
\Delta w_{\rm{com}}(t_p) & = & -\eta(\kappa + w_{\rm{com}}(t_p)) \label{B2} 
\end{eqnarray}
respectively.
Similarly, in case of failure ($r=0$) these changes are
\begin{eqnarray}
\Delta w_{\rm{win}}(t_q) & = & \eta(\kappa - w_{\rm{win}}(t_q)) -\rho + \varphi 
\label{B3} \\
\Delta w_{\rm{com}}(t_q) & = & - \eta(\kappa + w_{\rm{com}}(t_q)) + 
\varphi \label{B4}
\end{eqnarray}
respectively.
Only if the increase of $w_{\rm{com}}$ is larger 
than the increase of 
$w_{\rm{win}}$, the memory path can be wiped out, since then $w_{\rm{com}}$
may become the largest weight, i.e., if
\begin{equation}
\Delta w_{\rm{com}}(t_p) + \Delta w_{\rm{com}}(t_q) > 
\Delta w_{\rm{win}}(t_p) + \Delta w_{\rm{win}}(t_q) \label{B5}
\end{equation}
We now substitute (\ref{B1})--(\ref{B4}) into (\ref{B5}). 
In the resulting inequality we can ignore the values of $w_{\rm{win}}$ and 
$w_{\rm{com}}$ relative to 
$\kappa$ as long as the number of adaptations of $w_{\rm{win}}$ and 
$w_{\rm{com}}$ is
small; recall that $\kappa=1$, $\rho=0.02$, and the initial values of the 
weights are in the range $[-0.01, 0.01]$ in our numerical experiments. 
With these approximations, the inequality reduces to
$\rho > 4  \eta$. In the opposite case, 
\begin{equation}
\rho < 4\eta
\label{Bla1}
\end{equation}
$w_{\rm{win}}$ will remain larger than $w_{\rm{com}}$ and, consequently, path
interference will not wipe out learned input--output relations, which
explains the decrease of the number of learning steps for 
$\eta > \frac{1}{4}\rho$. 
For $\eta > \frac{1}{2}$ the network is incapable of learning input--output
relations. This can be seen as follows. Each time a winning connection is 
punished, i.e., the output is wrong, it changes approximately by an amount
$\eta-\rho$, whereas the competing connection changes by an amount of
$-\eta$. Hence, only when $\eta - \rho < -\eta$, or, equivalently, when
\begin{equation}
2\eta < \rho
\label{Bla2}
\end{equation}
the winning connection
decreases more than the competing connection. In the opposite case, 
$2\eta > \rho$, the winning connection remains larger than its competitor,
and, at the next learning step, the output will be wrong again.

Combining the inequalities (\ref{Bla1}) and (\ref{Bla2}), we find the 
central result of this article (\ref{ineq}), the parameter region for which 
a Hebbian term in the learning rule is advantageous. This observation is 
confirmed by the numerical experiment of section \ref{subsec:actexp}, so,
in particular, figure \ref{fig:vareta}.

Note that the reasoning leading to the main results (\ref{Bla1}) and 
(\ref{Bla2}) was based on an assumption regarding the initial values.
In particular, it was assumed that the weights were small compared to 
$\kappa$ (which was put equal to $1$). In reference \cite{He99} it was shown
that the pre-factor (\ref{eq:vareps1}) of the Hebbian term tends to zero during
the learning process:
\begin{equation}
\kappa - (2x_i-1)(h_i - \theta_i) \rightarrow 0
\end{equation}
implying, that for a small number of active neurons the absolute values of the
weights are of the order $\kappa$, as follows with (\ref{membranepotential}) 
and (\ref{thetazero}). Hence, the assumption that the weights are small
compared to $\kappa$ ($\kappa = 1$) is guaranteed to break down at a certain
point in the learning process.

\section{Summary}\label{sec:sum}

We have shown, in a particular model, that a Hebbian component in a 
reinforcement rule improves the ability to learn input--output relations
by a neural net.

\end{document}